\documentclass{article}

\usepackage{amsbsy} 
\usepackage{booktabs}	
	\usepackage{multirow} 
	\usepackage{multicol}

\usepackage{multicol}
\usepackage{graphicx}
\usepackage[version=3]{mhchem} 
\usepackage{setspace}
\usepackage{subfigure}
\usepackage{url}
\usepackage{xspace}
\usepackage{lastpage} 
\usepackage[left=3cm,bottom=3cm,top=3cm,right=3cm]{geometry}

\usepackage{tabularx}
\usepackage{array}
\usepackage[plainpages=false,
colorlinks,
debug,
linkcolor=red,
citecolor=red, 
filecolor=blue,
breaklinks=true]
{hyperref}

\usepackage{fancyhdr}
\fancypagestyle{plain}{ %
\fancyhf{} 
\setlength{\headheight}{15.2pt}
\setlength{\footskip}{40pt}
\lhead{\textit{The effect of \betap precipitates on a Mg-Zn alloy}}
\rhead{J.M.Rosalie\ \& H. Somekawa \& A. Singh \& T. Mukai}
\lfoot{\textit{Accepted author manuscript}}
\rfoot{Page \thepage\ of\ \pageref{LastPage}}

}

\newcommand{\betap}{\ensuremath{{\beta_1^\prime}}\xspace}
\newcommand{\micron}{\ensuremath{\mathrm{\mu m}}\xspace}	
\newcommand{\celsius}{\ensuremath{\mathrm{^\circ C}}\xspace} 

\newcommand{\fig}[2][6cm]{\resizebox{#1}{!}{\includegraphics{#2}}} 

\begin{document}


\title{The effect of size and distribution of  rod-shaped \betap precipitates on the strength and ductility of a Mg-Zn alloy}

\author{Julian~M~Rosalie$^a$ 
\and Hidetoshi Somekawa$^a$
\and  Alok Singh$^a$
\and Toshiji Mukai$^b$}
\date{}

\maketitle
\noindent$^a$ Structural Materials Unit, 
National Institute for Materials Science, 
Tsukuba, 305-0047, Japan.\\
$^b$ Dept. Mechanical Engineering, Kobe University, 1-1 Rokkodai, Nada, Kobe city, 657-8501 Japan.

\begin{abstract}  

We report on a quantitative investigation into the effect of size and distribution of rod-shaped \betap precipitates on strength and ductility of a Mg-Zn alloy.  
Despite precipitation strengthening being crucial for the practical application of magnesium alloys this study represents the first systematic examination of the effect of controlled deformation on the precipitate size distribution and the resulting strength and ductility of a magnesium alloy.
Pre-ageing deformation was used to obtain various distributions of rod-shaped \betap precipitates through heterogeneous nucleation. 
Alloys were extruded to obtain a texture so as to avoid formation of twins and thus to ensure that dislocations were the primary nucleation site. 
Pre-ageing strain refined precipitate length and diameter, with average length reduced from 440\,nm to 60\,nm and diameter from 14\,nm to 9\,nm.
Interparticle spacings were measured from micrographs and indicated some inhomogeneity in the precipitate distribution. 
The yield stress of the alloy increased from 273\,MPa to 309\,MPa.
The yield stress increased linearly as a function of reciprocal interparticle spacing, but at a lower rate than predicted for Orowan strengthening.
Pre-ageing deformation also resulted in a significant loss of ductility (from 17\% to 6\% elongation). 
Both true strain at failure and uniform elongation showed a linear relationship with particle spacing, in agreement with models for the accumulation of dislocations around non-deforming obstacles. 
Samples subjected to 3\% pre-ageing deformation showed a substantially increased ageing response compared to non-deformed material; however, additional deformation (to 5\% strain) resulted in only modest changes in precipitate distribution and mechanical properties. 

\end{abstract}





\section{Introduction}

Precipitation is one of the primary methods to improve the strength of magnesium alloys. 
Of the commercial Mg alloys the ZK series show the greatest precipitation strengthening  response \cite{ClarkMgZn1965}.
These alloys are based on the Mg-Zn binary system in which the strengthening precipitate is a high aspect-ratio rod, termed \betap, which adopts a \(\langle 0001 \rangle\) habit. 
These precipitates provide resistance to basal slip in magnesium \cite{ClarkMgZn1965} and  it has been suggested, even limit deformation twinning in alloys aged to optimal hardness \cite{Chun1969}. 

Precipitation strengthening is controlled by the interparticle spacing and is thus influenced by the precipitate size and number density.
For non-spherical precipitates the precipitate morphology, crystallographic habit and aspect ratio must also be considered. 
Rod-shaped morphology parallel to the hexagonal axis is more effective in inhibiting basal or prismatic slip than spherical precipitates or plate shaped precipitates in the basal plane, because a given volume fraction of precipitate intersects more basal planes \cite{NieMg2003, Robson2011}. 

Introducing lattice defects in Mg-Zn alloys has been shown to accelerate and enhance the ageing process \cite{MimaAgeing1971a}.
These defects provide heterogeneous nucleation sites and  precipitation of \betap on dislocations has been noted by various workers \cite{ClarkMgZn1965,Singh2007,Ohishi2008}. 
Trace alloying elements  have also been used to refine the size and increase the number density of the precipitates \cite{MendisTrace2007,Mendis2009,GengTrace2011}. 
However, despite the widespread recognition of the importance of precipitation strengthening in magnesium alloys, no quantitative investigation into the effect of size and distribution of the precipitates has yet been reported. 

Of equal importance to the strength is the inherently lower ductility of magnesium alloys, which is further reduced by precipitation strengthening. 
However, although Mg-Zn alloys form the basis of commercial ZK alloys as well as experimental alloys, the influence of \betap precipitates on the ductility of precipitation-hardened alloys has not been investigated in any detail.  

In this investigation the size and distribution of the rod-shaped \betap precipitates in a binary Mg-Zn alloy has been modified by introducing heterogeneous nucleation sites through deformation prior to ageing. 
To avoid complications arising out of twinning, we have used textured alloys obtained by means of extrusion, ensuring that deformation occurs near-exclusively via slip.
The effect of the modified distribution of the precipitates on tensile strength and ductility were studied. 

\section{Experimental details}

Billets of binary Mg-3.0at.\%Zn were prepared from pure elements via direct-chill casting, with the composition being confirmed via inductively-coupled plasma mass spectroscopy (ICP-MS). 
The billets were homogenised for 15\,h  at 300$^\circ$C and then extruded into 12\,mm diameter rods at an extrusion ratio of 12:1 and temperatures of 300$^\circ$C in order to develop strong texture. 
Cylindrical dog-bone samples (gauge length 15\,mm, diameter 3\,mm) were machined from the extruded rods.  
These were encapsulated in helium and solution treated for 1\,h at 300$^\circ$C for alloy and quenched into water at ambient temperature. 

An Instron mechanical tester was using to impose controlled amounts of pre-ageing deformation.
The test pieces were deformed in tension (i.e. strained parallel to the extrusion axis) at a strain rate of \(1\times10^{-3}\,\mbox{s}^{-1}\).  
Samples were deformed to a nominal plastic strain of either 3\% or 5\%, with additional sample retained in the non-deformed state for comparison.  
Samples were then aged to peak hardness 
in an oil bath at 150\celsius for periods of 256\,h for non-deformed samples and 32-48\,h for deformed samples.
The ageing response was measured using Vickers hardness testing with a 300\,g load. 
Tensile tests were carried out to failure on peak-aged samples using the Instron mechanical tester, at strain rates of \(1\times10^{-3}\,\mbox{s}^{-1}\). 
Solution-treated and quenched samples (8\,mm $\times$ 4\,mm diameter)  were also tested in compression to measure the tensile-compressive asymmetry. 

Samples for TEM analysis were prepared from the aged specimens by grinding to \(\sim70\,\micron\) and then thinning to perforation using a Gatan precision ion polishing system. 
TEM observations were conducted using a JEOL 4000EX instrument operating at 400\,kV. 
Measurements of precipitate length, diameter and interparticle spacing were made on scanned negatives using ImageJ software (version 1.44).

The cross-sectional areas of individual precipitates were directly determined from the images recorded along [0001] zone axis of the matrix grains by the analysis software (ImageJ). 
Precipitate diameters were taken as the diameter for a cylindrical rod of equivalent area. 
The average centre-centre distance between \betap particles was determined by Delaunay triangulation, using the Delaunay Voronoi plug-in for ImageJ.

\section{Results}

Optical microscopy examination confirmed that deformation twins were virtually absent from all samples, 
confirming that deformation had occurred almost exclusively via slip. 
In all deformation conditions volume fractions of only  1-2\% twins were present, suggesting this may be a processing artifact. 
A grain size of $28\pm3\,\mu$m was obtained from line-intercept measurements.

\subsection{Ageing response}

\begin{figure} [htpb] 
\begin{center}
\includegraphics{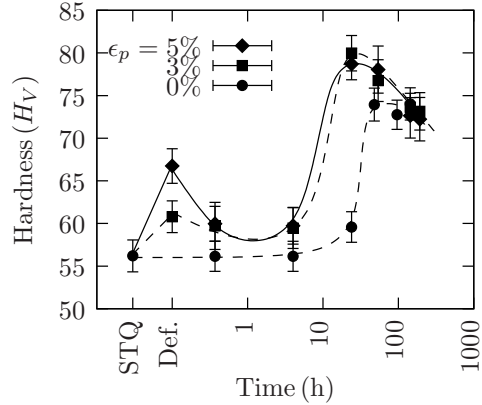}
\caption{Ageing response for Mg-Zn alloys at 150\celsius as a function of pre-ageing strain ($\epsilon_p$). ``STQ''  indicates the hardness in the solution treated and quenched condition, while ``Def.'' shows the hardness for the deformed samples prior to ageing. 
 \label{fig-hardness}}
\end{center}
\end{figure}

Deformation resulted in an immediate increase in the hardness with 5\% deformation raising the hardness to 66\,$H_V$ (Figure~\ref{fig-hardness}). 
Upon ageing,  this was followed by a rapid drop in hardness and a plateau region before the hardness increased to its optimum value. 
This indicates work-hardening during pre-ageing deformation, followed by annealing out of dislocations during the early stages of ageing. 
Deformed alloys showed a more rapid and extensive response to ageing; however, aside from the greater hardness prior to ageing, there was little difference between the behaviour of  the alloy when deformed to either 3\% or 5\% in terms of either peak hardness or optimum ageing time. 

\subsection{Precipitate size and distribution}

\begin{figure} [htpb] 
\begin{center}
	\hfill
\subfigure[0\% strain
	\label{fig-3000Z-0-side}]{\fig[0.37\textwidth]{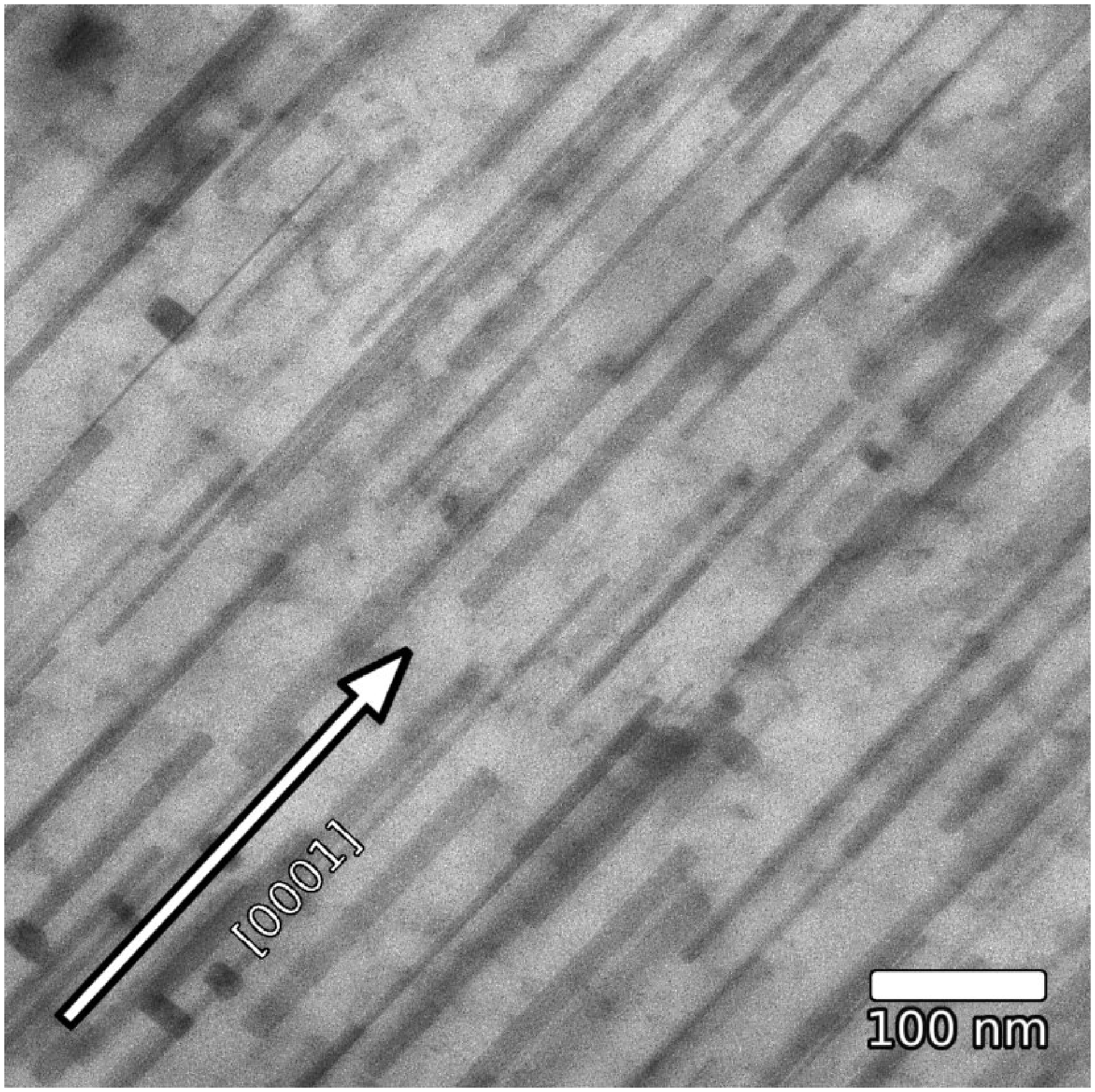}} \hfill
\subfigure[3\% strain 
	\label{fig-3000Z-3-side}]{\fig[0.37\textwidth]{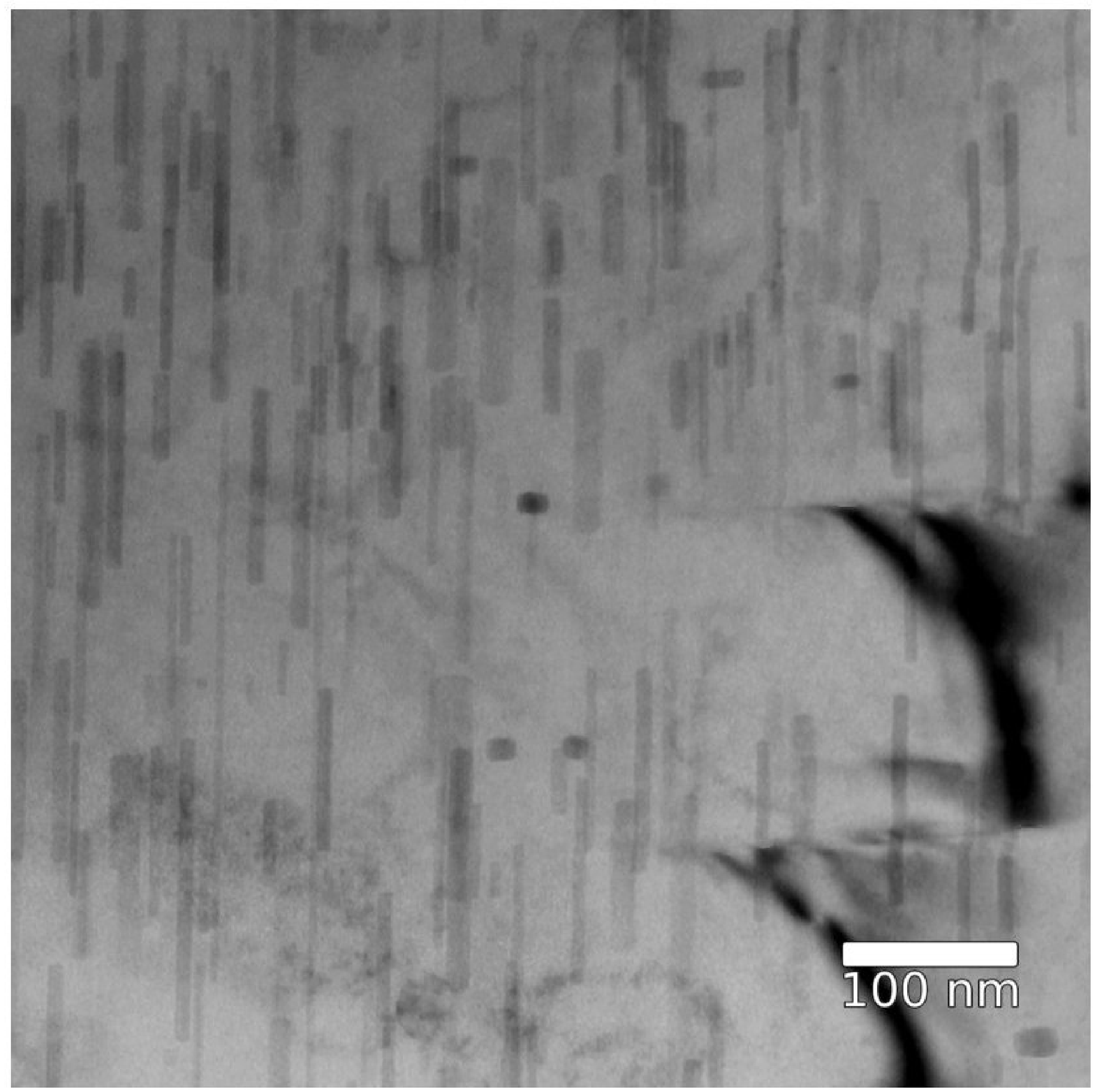}} \hfill\

\subfigure[5\% strain 
	\label{fig-3000Z-5-side}]{\fig[0.37\textwidth]{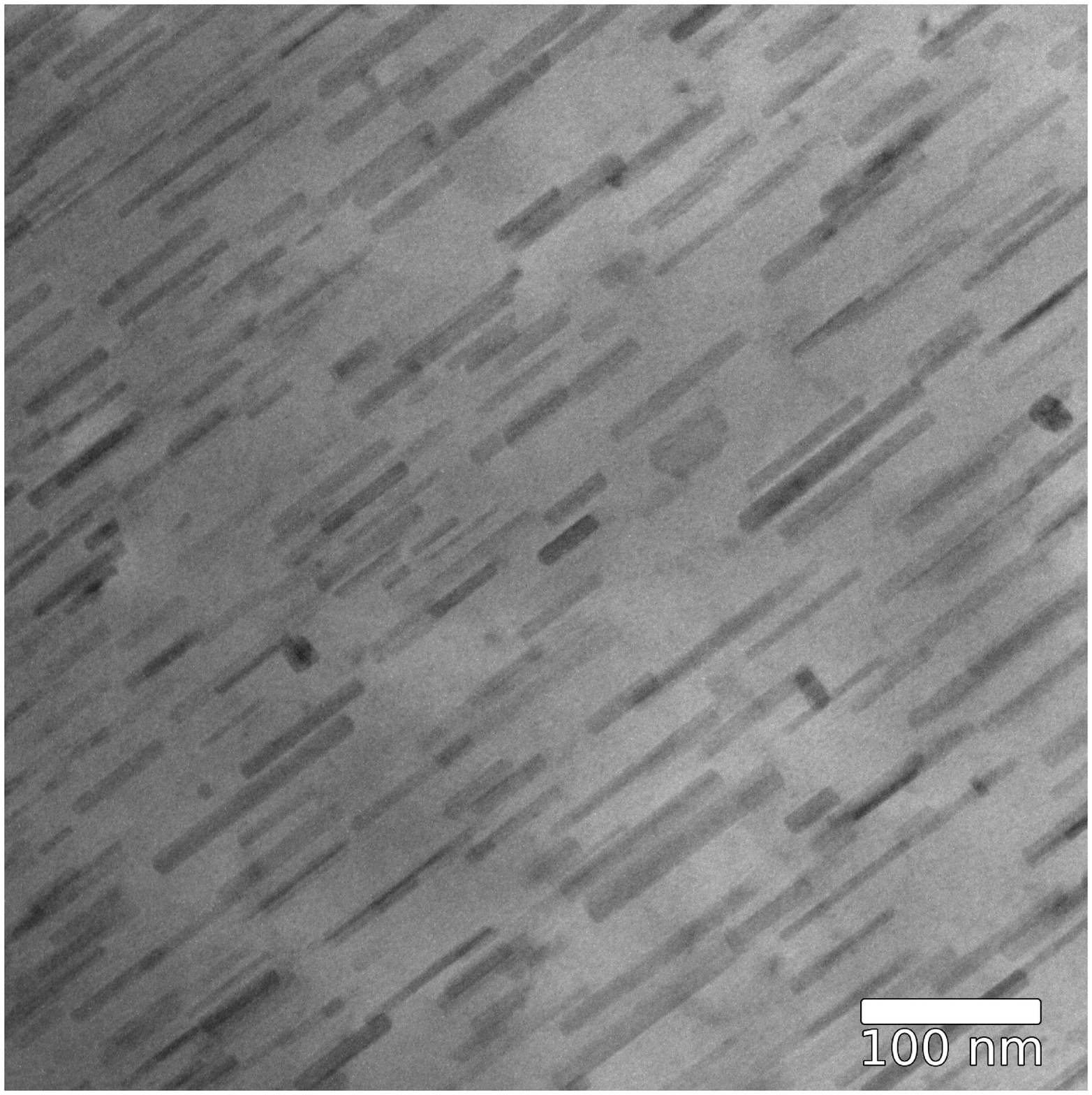}} 
\caption{Transmission electron micrographs of \betap precipitates in the optimum hardness condition as a function of pre-ageing deformation. 
The electron beam was directed normal to [0001]$_\textrm{Mg}$.
 \label{fig-tem-side}}
\end{center}
\end{figure}

The principle intragranular precipitate present at peak age were rod-shaped \betap precipitates aligned parallel to the [0001] axis of the Mg matrix. 
Figure~\ref{fig-tem-side} shows representative micrographs of the precipitates obtained with the electron beam normal to the [0001] axis.

\begin{figure} [htpb] 
\begin{center}
\includegraphics{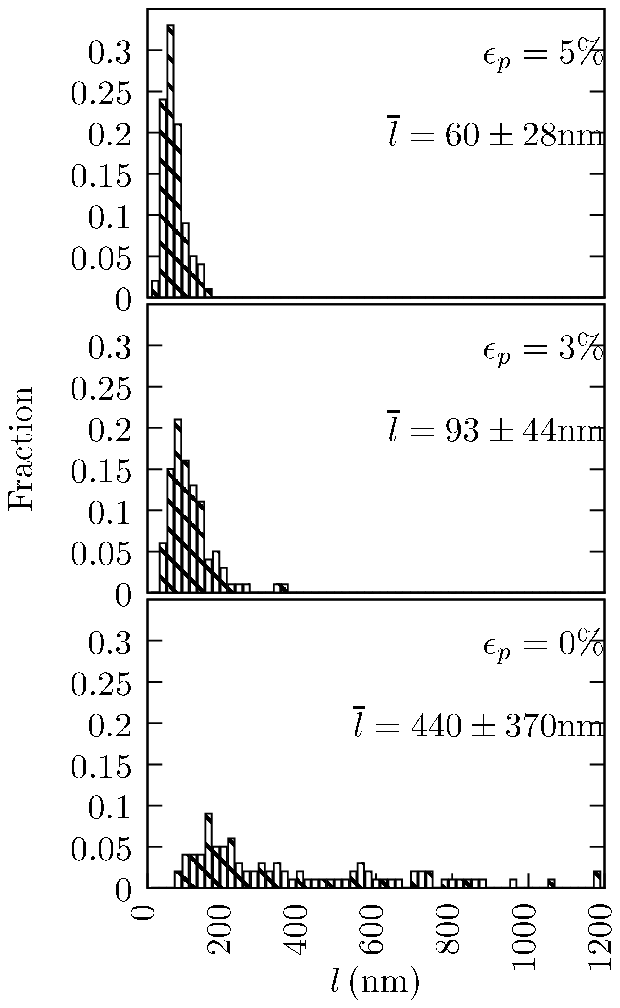}
\caption{Precipitate length, $l$, for the rod-like \betap precipitates in the peak aged condition as a function of pre-ageing deformation ($\epsilon_p$).  
\label{fig-length}}
\end{center}
\end{figure}

The distribution of precipitate length values is provided in Figure~\ref{fig-length} and shows that 
pre-ageing deformation resulted in a reduction of the length of the \betap precipitates. 
Samples deformed before ageing showed narrower precipitate length distributions with the average \betap precipitate length  decreasing from 440\,nm for 0\% strain to 60\,nm for 5\% strain. 

\begin{figure} [htpb] 
\begin{center}
\hfill
\subfigure[0\% strain 
\label{fig-3000Z-0-cross}]{\fig[0.37\textwidth]{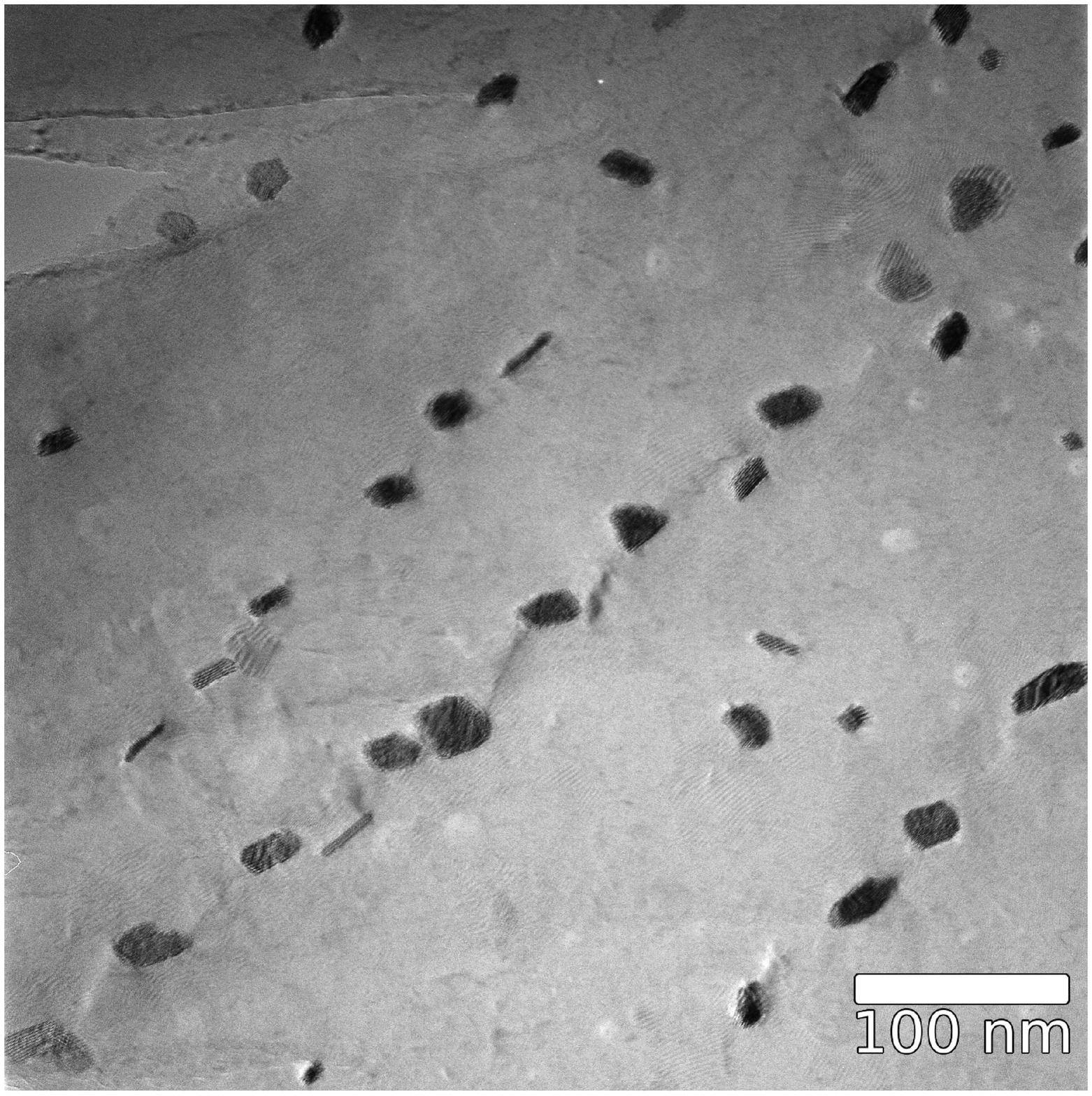}} 
\hfill
\subfigure[3\% strain 
\label{fig-3000Z-3-cross}]{\fig[0.37\textwidth]{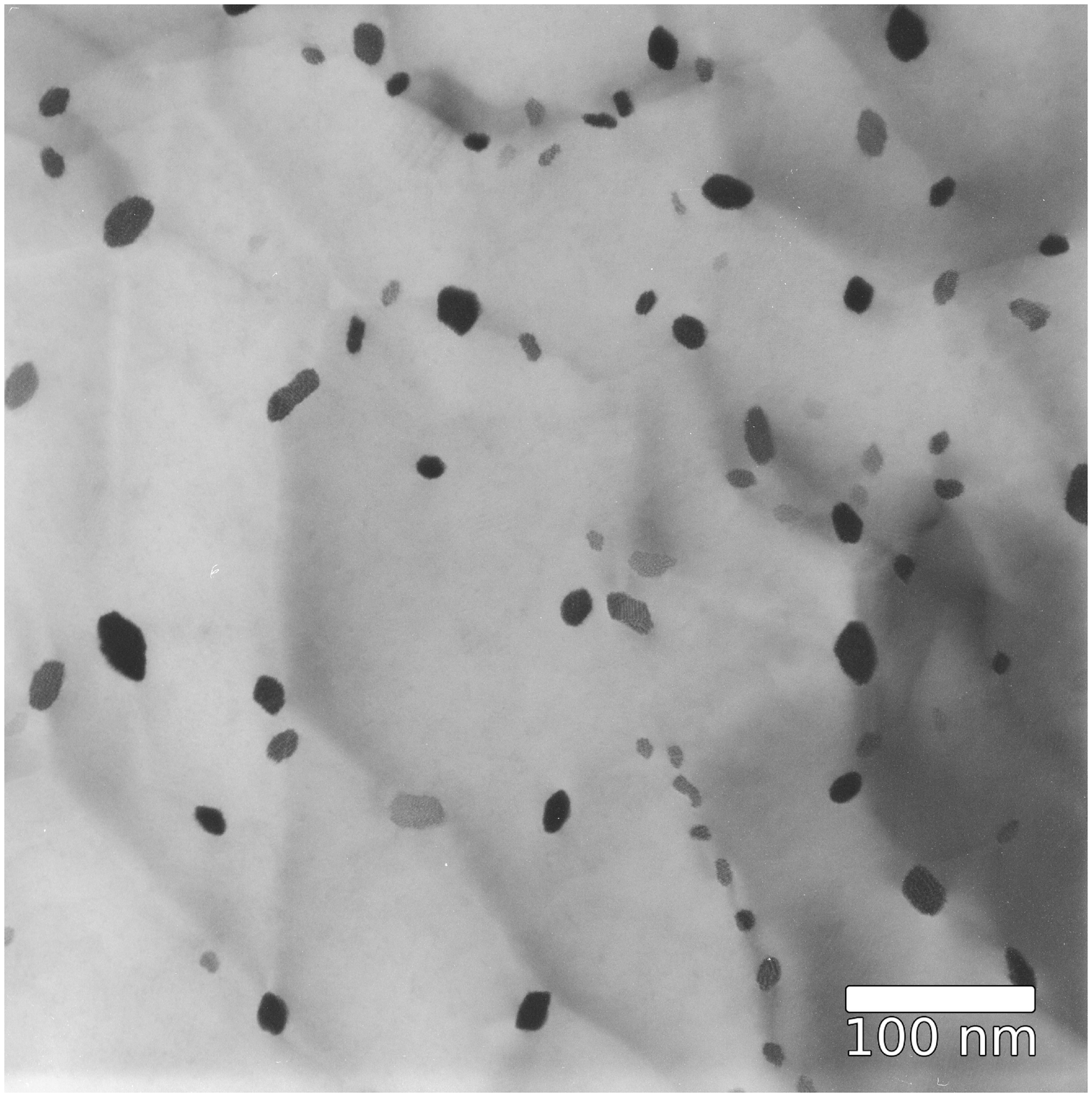}}\hfill\

\subfigure[5\% strain 
\label{fig-3000Z-5-cross}]{\fig[0.37\textwidth]{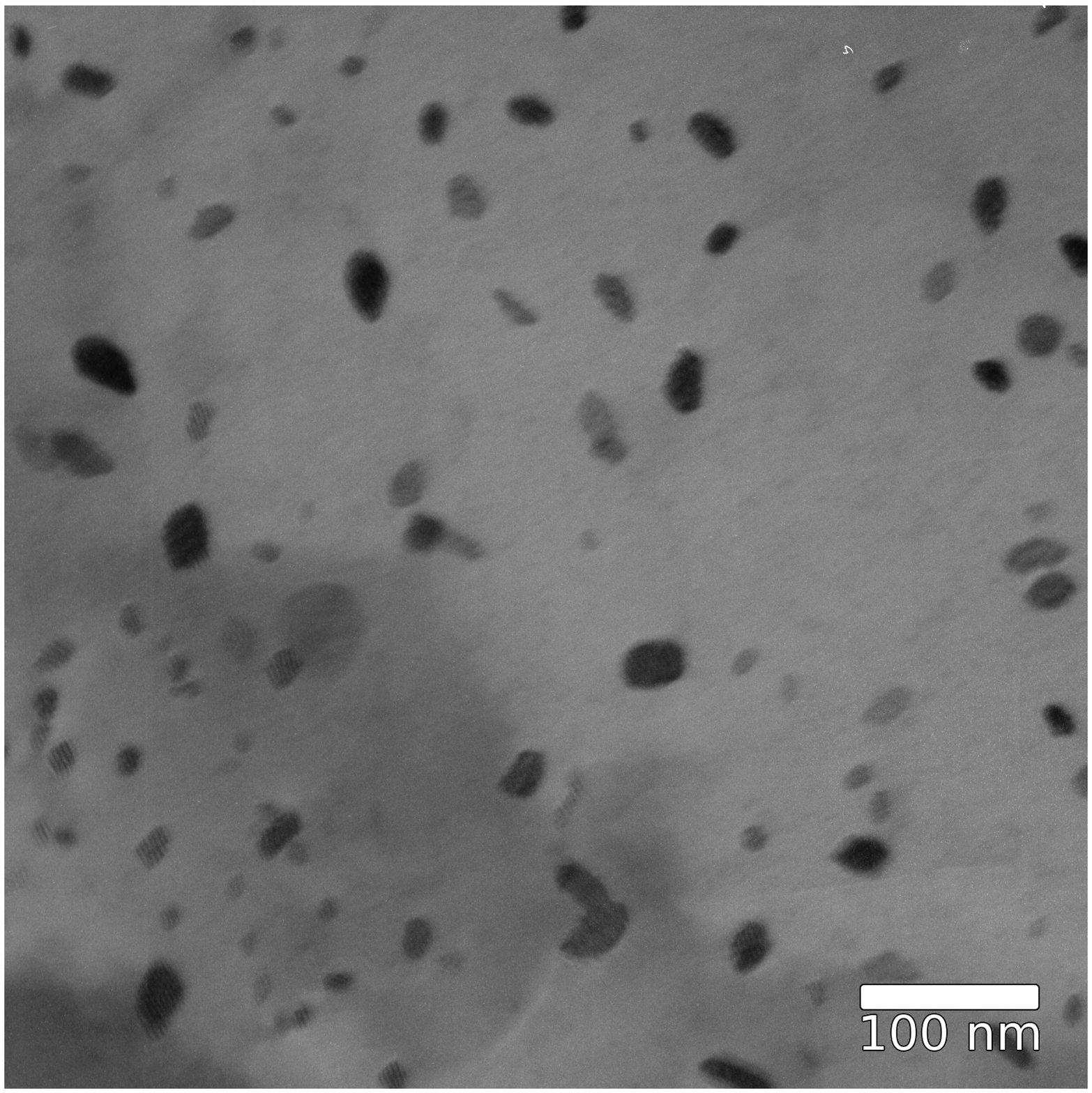}}\hfill
\caption{Transmission electron micrographs of \betap precipitates in the optimum hardness condition as a function of pre-ageing deformation. 
The electron beam was directed parallel to [0001]$_\textrm{Mg}$.
\label{fig-tem-cross}}
\end{center}
\end{figure}

\begin{figure} [htpb] 
\begin{center}
\includegraphics{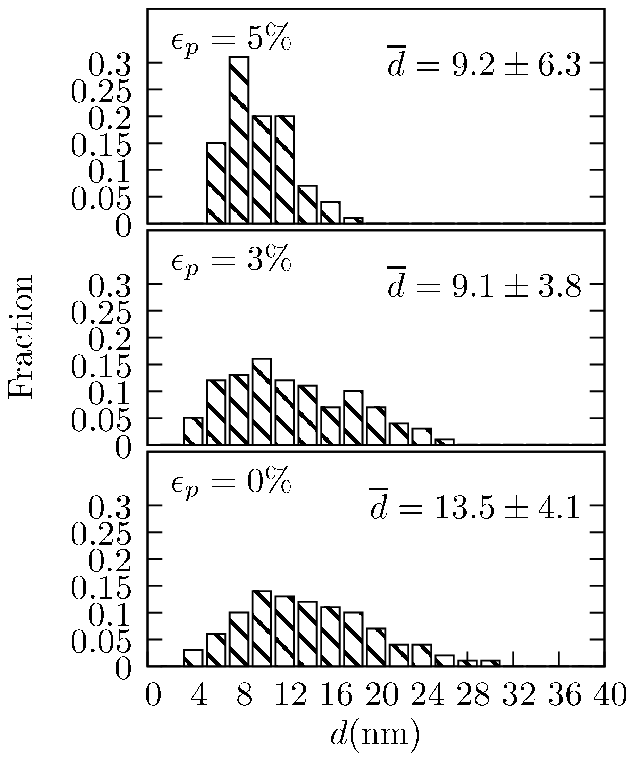}
\caption{Precipitate diameter, $d$, for the rod-like \betap precipitates at peak age as a function of pre-ageing deformation. \label{fig-diameter}}
\end{center}
\end{figure}

Pre-ageing deformation also affected the precipitate diameter.
Precipitates with similar thicknesses to those in non-deformed sample were observed; in addition to which a number of thinner precipitates occurred.  
The \betap precipitate distribution in the non-deformed peak-aged alloys is considerably sparser than in the deformed samples. 
The \betap precipitates showed a tendency to form along the lines of strain contrast which indicated dislocations.
Figure~\ref{fig-tem-cross}  shows micrographs for each deformation condition with the beam directed along the [0001] axis to show the precipitates in cross section. 
The precipitate diameter distribution is given in Figure~\ref{fig-diameter}. 

The extend of inhomogeneity in the precipitate distribution was evaluated by comparing interparticle spacings measured on the basal plane with those calculated for a uniform distribution.
The interparticle spacings on the basal plane were  measured directly from the micrographs by subtracting the average average particle diameter  from the average centre-centre distance obtained via Delaunay triangulation \cite{Lepinoux2000}.
Interparticle spacings for the alloys in all deformation conditions are set out in Table~ \ref{tab-lambda}. 
The precipitate diameter and interparticle spacing were measured in thin regions close to the edges of the foil to minimise potential overlap. 

\subsection{Tensile Strength and Ductility}

\begin{figure} [htpb] 
	\begin{center}
\includegraphics{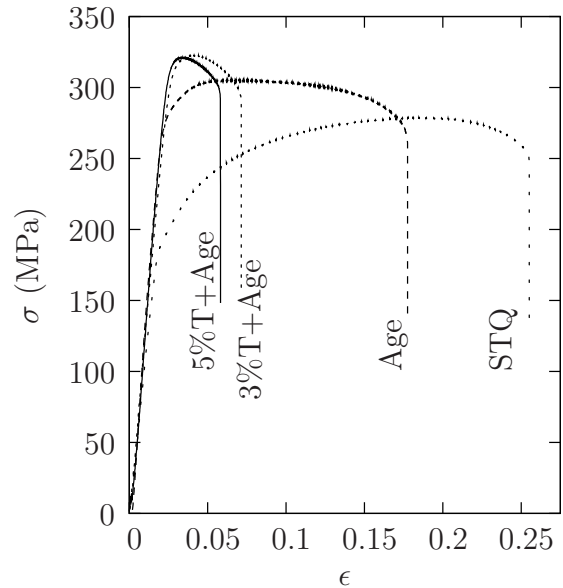}
	\caption{\label{fig-instron2} Engineering stress ($\sigma$)- strain($\epsilon)$ curves for  Mg-Zn as a function of nominal pre-ageing deformation (T\%).}
	\end{center}
\end{figure}

Tensile samples were tested to failure in (a) the solution-treated condition, (b) after peak ageing and (c) after 3\% or 5\% strain and peak ageing.
Stress-true strain curves are provided in Figure~\ref{fig-instron2} and show a substantial increase in the yield strength upon ageing, accompanied by a decrease in the elongation to failure. 
The key data is set out in Table~\ref{tab-mechanical}.

\begin{table}
\begin{center}
\caption{Mechanical behaviour as a function of thermo-mechanical treatment.
Yield strength ($\sigma_y$), ultimate tensile strength (UTS) and failure strain ($\epsilon_f$) were determined from the strain- tensile curves. 
The true stress at failure ($\epsilon_{fr}$) was calculated from the area reduction at the point of fracture. 
Values in parentheses indicate the standard error in the last digit as a measure of uncertainty. \label{tab-mechanical}}
\begin{tabularx}{0.45\textwidth}
{l@{ }
>{\raggedleft}X@{(}c@{)}
>{\raggedleft}X@{(}c@{) }
>{\raggedleft}X@{(}c@{) }
>{\raggedleft}X@{(}c@{)}}\toprule
& \multicolumn{2}{c}{$\sigma_y$} &  \multicolumn{2}{l}{UTS} & \multicolumn{2}{c}{$\epsilon_f$}  &\multicolumn{2}{c}{$\epsilon_{fr}$} \\
	 & \multicolumn{4}{c}{(MPa)} \\	\cmidrule(r){2-9}
STQ	&	143	&	5	&	273	&	5	&	0.24	&	0	&	0.25	&	1	\\
0\%	&	273	&	1	&	305	&	1	&	0.17	&	0	&	0.16	&	1	\\
3\% .	&	305	&	2	&	322	&	1	&	0.06	&	1	&	0.15	&	2\\
5\%.	&	309	&	6	&	323	&	3	&	0.06	&	0	&	0.15	&	1\\
\bottomrule 
\end{tabularx}
\end{center}
\end{table}

Solution-treated samples 
had 0.2\% yield strengths of $143\pm5$\,MPa in tension. 
The compressive yield strength of the solution-treated material was considerably lower, $107\pm3$\,MPa,  as expected for an extruded Mg alloy with strong texture.
Isothermal ageing substantially increased the proof strength in tension of the binary alloy to 273$\pm1$\,MPa. 
Pre-ageing deformation resulted in further increases in the proof strength, to about $304\pm2$\,MPa.
The difference in proof stress between 3\% and 5\% deformations was minor, with 5\% pre-strain increasing the yield strength to $309\pm5$\,MPa.
It should also be noted that for the deformed samples these values were close to the ultimate tensile strengths of $321\pm1$\,MPa and $322\pm2$\,MPa, respectively.

\section{Discussion}
\subsection{Interparticle spacings}

For a magnesium alloy extruded and deformed in tension, slip is expected to  occur predominantly on prismatic planes with no widespread pyramidal $\langle c+a \rangle $  slip  \cite{AlSamman2010}.
This is consistent with the much greater stress required to initiate $\langle c+a \rangle $ slip \cite{Raeisinia2010}.
However, Koike \textit{et. al}\cite{KoikeNonBasalSlip}, found that significant cross-slip between basal and non-basal planes occurred even at room temperature. 
Since prismatic and basal slip share identical Burgers vectors, the  potential for both basal and prismatic slip was considered and interparticle spacings were determined for both slip systems. 

\subsubsection{Basal slip}

Interparticle spacings are generally calculated from  using standard stereological  relationships, which assume a homogeneous distribution of particles. 
However, the TEM micrographs (Fig.~\ref{fig-tem-cross}) showed a strong tendency for \betap precipitates to form along dislocations (as noted in in previous reports \cite{ClarkMgZn1965,Singh2007,Ohishi2008}).
As this suggesting that the precipitates may not be uniformly distributed throughout the matrix the average spacing on the basal plane was determined from micrographs via Delaunay triangulation. 

For a homogeneous distribution of rod-like precipitates with $\langle 0001 \rangle$ habit impeding basal slip in Mg the spacing, $\lambda$, is given by \cite{NieMg2003}: 
\begin{equation}
 \lambda = \left( \frac{0.953}{\sqrt{f}} -1 \right) d_1 
\label{eq-lambda}
\end{equation}

\begin{figure} [htpb] 
\begin{center}
\fig[0.48\textwidth]{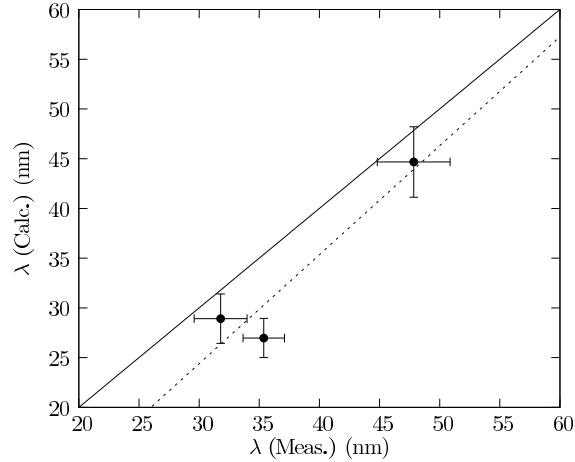}
\caption{Measured vs. calculated interparticle spacings on the basal plane ($\lambda$). \label{lambdacompare}}
\end{center}
\end{figure}

Figure~\ref{lambdacompare} compares the measured spacing of the \betap precipitates with values calculated from Equation~\ref{eq-lambda}.
Since the alloys were in the peak-aged condition it was assumed that the level of zinc in solution was the equilibrium value at the ageing temperature (0.07at\%) with the remaining zinc partitioned to \betap precipitates. 
For \betap precipitates with a composition \ce{Mg4Zn7} and a density of 4.8\,g\,cm$^{-3}$ \cite{Yarmolyuk1975} alloy this gives a volume fraction of 3.5\%\footnote{It has been shown that the precipitates contain domains of \ce{MgZn2} (density 5.0\,g\,cm$^{-3}$\cite{Komura1972}) however even assuming the particle composition as \ce{MgZn2} the estimated volume fraction is 3.2\%.}.
The solid line indicates a homogeneous distribution (i.e. $\lambda$(measured) =$\lambda$(calculated)), with the measured values being greater than from the stereological calculations. 
Although the data is somewhat scattered, the measured values are approximately 15\% greater than for a homogeneous particle distribution. 
The measured particle spacings were used in further calculations of the strength and ductility.

\subsubsection{Prismatic slip}

Robson \textit{et al.} determined the spacing for prismatic slip around rod-like precipitates with $\langle 0001 \rangle$ habit assuming a square array of particles and found a spacing of: \cite{Robson2011}: 
\begin{equation}
 \lambda = \frac{1}{\sqrt{N_A}} - l_1 
\label{eq-lambda}
\end{equation}
with the number density on a single slip plane given by ; $N_A=N_V d_t$ where $N_V$ is the number density per unit volume and $l$ is the precipitate diameter.

\begin{figure} [htpb] 
\begin{center}
\subfigure[\label{Triangle}]{\includegraphics[width=.45\textwidth]{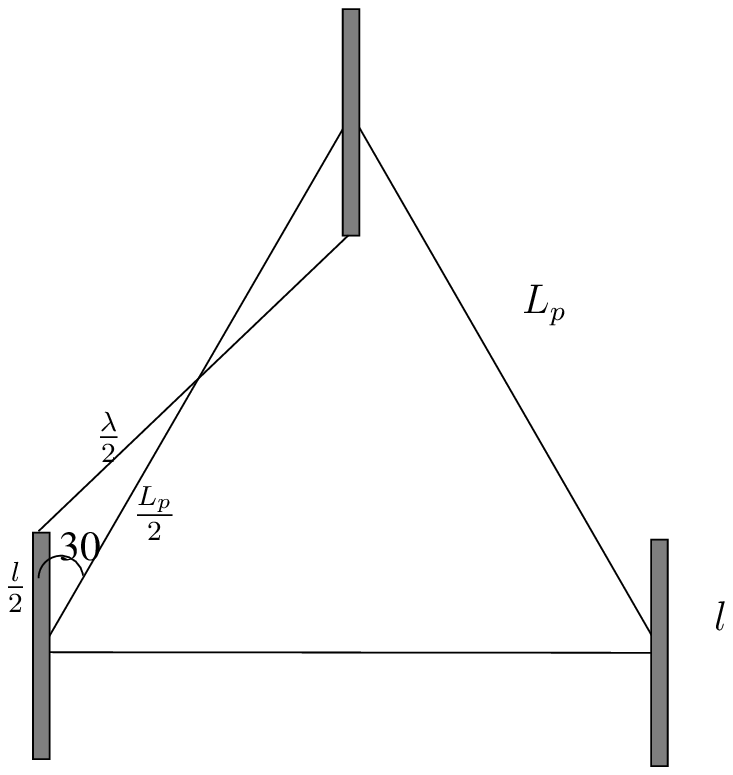}}
\subfigure[\label{Tri-graph}]{\includegraphics[width=.45\textwidth]{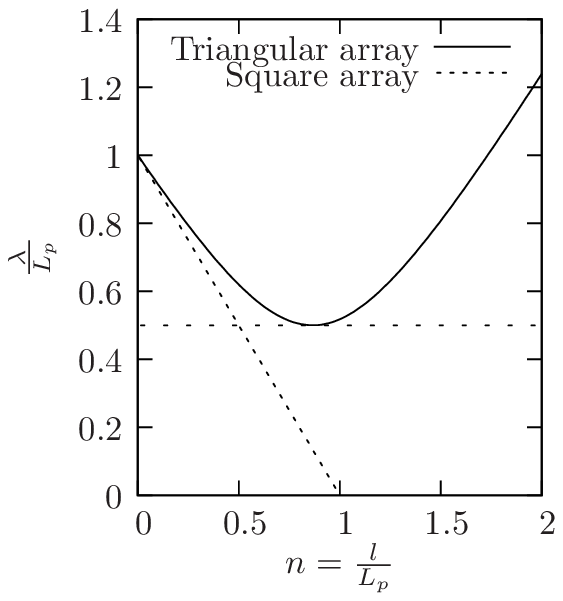}}
\caption{a) Schematics of the triangular array of particles used to calculate interparticle spacings on the prismatic plane and b) the interparticle spacing as a function of the length of particles/the centre-centre distance.}
\end{center}
\end{figure}

The choice of a square array of particles considerably simplifies the mathematics, however this arrangement is highly sensitive to the aspect ratio and might be problematic for particles of the order of 450\,nm with aspect ratios 15-20 of as found in this work. 
It was deemed unlikely that two rods would be formed in perfect alignment and so in this work the spacing was calculated using a triangular array (See Figure~\ref{Triangle}).
If the particle length $(l)$ is expressed a a multiple $(n)$ of the centre-centre distance $(L_p)$,
then from the cosine formula:
\begin{equation*}
(\lambda/2)^2 	  =  (l/2)^2 + (L_p/2)^2 - 2 l L_p \cos 30^\circ \\
\end{equation*}
which yields
\begin{equation*}
\lambda			 =  L_p \sqrt{ \left( (n^2+1 - n\sqrt{3} \right)}\\
\end{equation*}
and 	since $L_P = 1/\sqrt{N_A}$
\begin{equation}
\lambda			 =  \frac{1}{\sqrt{N_A}} \sqrt{ \left( (n^2+1 - n\sqrt{3} \right)} \label{pyramid}
\end{equation}

The relative spacing  $(\frac{\lambda}{L_p})$ as a function of $n$ is plotted in Figure~\ref{Tri-graph} for both square and triangular arrays.
While the difference between the two models is negligible for $l \ll L_p$ the curves diverge rapidly as $n\rightarrow 1 $.
It was also noted that for a triangular array the spacing increases in a physically unrealistic way for $n>1$ and therefore a lower limit of $(\frac{\lambda}{L_p})=0.5$ (the minimum of the curve) was imposed.

Interparticle spacings for pyramidal plane (as calculated by Equation~\ref{pyramid}) are listed in Table~\ref{tab-lambda-py}. 
The centre-centre distances were determined from precipitate length and diameter values and it can be seen that the precipitate length is greater than half the interparticle spacing in each case.
Therefore, as described above, the effective interparticle spacing ($\lambda_{eff}$) was taken as $0.5L_p$.

\begin{table}
\begin{center}
\caption{Stereological measurements for \betap precipitates in alloys aged to optimum hardness as a function of pre-ageing  deformation. Values in parentheses indicate the standard deviation as a measure of the spread within the distribution. 
\label{tab-lambda}}
\begin{tabular}{p{5ex}p{7ex}p{7ex}p{7ex}} \toprule
Strain \newline $\epsilon_p$ \newline \%&  Prec. diam.   $d$ (nm) &Prec. len.~$l$ (nm)& Spacing (basal) \newline $\lambda_b$ \newline (nm) \\
 \midrule
  0    &   14(1)  & 440(50) & 48(4) \\
  3    &   9(0.4)  & 102(6) & 35(2)   \\
  5    &   9(1)  & 60(4) & 32(4)  \\ \bottomrule
\end{tabular}
\end{center}
\end{table}

\begin{table}
\begin{center}
\caption{Stereological measurements for prismatic slip around \betap precipitates in alloys aged to optimum hardness as a function of pre-ageing  deformation. Values in parentheses indicate the standard deviation as a measure of the spread within the distribution. 
\label{tab-lambda-py}}
\begin{tabular}{p{5ex}p{7ex}p{7ex}p{7ex}p{7ex}p{7ex}} \toprule
Strain \newline $\epsilon_p$ \newline \%&  Centre-centre \newline $L_p$ \newline (nm) &Prec. length (nm)&  $n$ & Spacing  \newline $\lambda_p$ \newline (nm) & $\lambda_{eff}$ \newline (nm) \\
 \midrule
  0    &   309 & 440(50) & 1.42 & 231 & 154\\
  3    &   107  & 102(6) &  0.87  & 54 & 54\\
  5    &   91  & 60(4) & 0.66  & 49 & 46\\ \bottomrule
\end{tabular}
\end{center}
\end{table}

\subsection{Strengthening}

The  increase in yield strength ($\Delta \sigma$) for a given volume fraction of homogeneously distributed, non-shearable  particles is given as \cite{NieMg2003}: 

\begin{equation}
\label{eq-mg-orowan-rod}
\Delta\sigma =
 \frac{Gb}{2\pi\sqrt{1-\nu}} 
\frac{1}{\lambda}
\ln{\frac{d_1}{\boldsymbol{b}}} 
\end{equation}
\noindent where $\nu$ is Poisson's ratio, $G$	shear modulus/GPa and $\boldsymbol{b}$ is the magnitude of the Burgers vector for $\langle a \rangle$ dislocations in Mg. 
(This Burgers vector is identical for both basal and prismatic slip.)
Values of G=16.6\,GPa, $\boldsymbol{b}$=0.32\,nm  \cite{AvedesianMg99} and $\nu$=0.28 were used in the calculations.
The planar diameter $(d_1)$ value for prismatic slip around rod-like particles will be the precipitate diameter in basal slip and precipitate length for prismatic slip.
From Equation~\ref{eq-mg-orowan-rod} the increment as a function of the reciprocal particle spacing will be: 
\begin{equation}
\frac{\Delta\sigma}{d (1/\lambda)} =
 \frac{Gb}{2\pi\sqrt{1-\nu}}
\ln{\frac{d_1}{\boldsymbol{b}}} 
\end{equation}

\begin{figure} [htpb] 
\begin{center}
\includegraphics{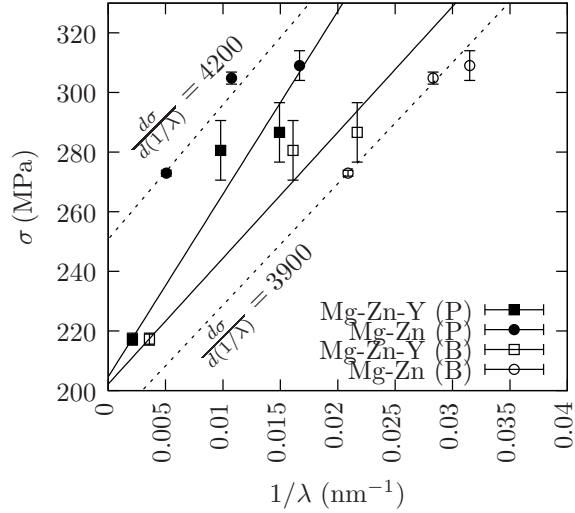}
\caption{Increments in strength ($\sigma$) plotted against reciprocal of precipitate spacing 
($1/\lambda$) 
in the prismatic (filled circles) and basal (open circles) planes   \label{sigma}}
\end{center}
\end{figure}

Figure~\ref{sigma} plots the yield strength of the peak-aged alloys  against the reciprocal of the calculated interparticle spacing (1/$\lambda$) on the prismatic and basal planes. 
Filled and open circles indicate 1/$\lambda$ values for prismatic and basal slip, respectively (see Tables \ref{tab-lambda} and \ref{tab-lambda-py}).
Both graphs show broadly linear increases in yield strength with increasing reciprocal spacing, however, extrapolation to $1/\lambda$=0 gives a much greater yield strength for prismatic slip than basal (262\,MPa $vs.$ 200\,MPa).
The latter value is substantially greater than the strength of the alloy in the solution treated condition and the intercept calculated for basal slip appears more representative of the alloy.

For prismatic slip with  a $d_1$  value in the mid-range of the experimental data, the equation predicts a gradient of 4760 MPa.nm. 
Least-squares fitting through the data in Figure~\ref{sigma} give gradients of only 3550\,MPa.nm for basal slip and 3100\,MPa.nm for prismatic slip.
Given that the yield strengths of the 3\% and 5\% strained samples are separated by only 5\,MPa any errors in the gradient will be magnified, however these lower-than-expected values would suggest that the particles are less effective obstacles to slip than if Orowan looping on either basal or prismatic planes was the only active deformation mechanism.

It is thought that cross-slip between basal and prismatic planes may account for the lower than expected increase in yield strength with reciprocal interparticle spacing.
While most work has focused on cross-slip at high temperatures (e.g.\cite{LukacMagnesium2011,Mathis2004}), Koike \textit{et al.} observed significant levels of cross-slip between basal and non-basal planes at room temperature in an AZ31B alloy \cite{KoikeNonBasalSlip}.
It was reported that this was due to plastic incompatibility stresses at the grain boundaries in the alloy. 
The present alloy had a considerably larger grain size ($28\pm3\mu$m \textit{vs.} $6.5\pm0.4\mu$m) and the influence of grain boundaries is likely to be much less pronounced.
However, the dense distribution of  non-shearable \betap particles should also present obstacles capable of inducing cross-slip between basal and prismatic planes, allowing dislocation with screw character to continue to glide through the matrix. 

\subsection{Ductility}

Previous studies of the effect of precipitation on ductility in aluminium alloys and steels have modelled the reduction in ductility by considering the the accumulation of geometrically necessary dislocations due to the difference in elastic moduli between the precipitates and the matrix. 
For non-shearable precipitates such as \betap rods the geometric slip distance in aged alloys is effectively the interparticle spacing, $\lambda$. 
The dislocation density ($\rho$)  is given by:
$\rho = (4 \epsilon /\lambda\mathbf{b})  $
where $\epsilon$ is the  strain and $\mathbf{b}$ is the Burgers vector for slip \cite{Ashby1970}. 

Chan \cite{chan1995} and Liu \textit{et. al} \cite{LiuZhang2004} modelled ductility in precipitate-strengthened materials  by
 assuming that failure occurs when the local dislocation density reaches a critical value $\rho_{cr}$.
The local critical stress at which this occurs, $\epsilon_{cr} $ is then 
\begin{equation}
\epsilon_{cr} = \frac{1}{4}\rho_{cr} \mathbf{b} \lambda \label{eqn-critical-stress}
\end{equation}

The local strain in the vicinity of the precipitates will be greater than the macroscopic strain. 
The macroscopic strain to failure, $\epsilon_f$,  cylindrical rod precipitate of length $l$ can be determined from the equation:
\begin{equation}
\epsilon_f = 
\frac{1}{\tilde{\epsilon}_E(\theta)} 
 \left[
\frac{I}{0.405 \pi h }
\right]^ \frac{1}{n+1}
\frac{\tilde{\epsilon_{cr}}}{2} 
\end{equation}
\noindent where $\tilde{\epsilon}_E(\theta)$ is a constant  and the co-efficients $n$ is the strain-hardening behaviour of the base material and 
\begin{eqnarray}
I &= &10.3 \sqrt{0.13+n}-4.8n\\
h &= &\frac{3}{2\sqrt{1+3n}}
\end{eqnarray}
\cite{LiuSun2005,Dowling1987}

Substitution into Equation~\ref{eqn-critical-stress} provides an expression for the failure strain in terms of $(\lambda)$, precipitate length, $l$, and the Ramberg-Osgood work hardening co-efficient of the matrix , $n$. 

\begin{equation}
\epsilon_f   = 
\frac{1}{\tilde{\epsilon}_E(\theta)} 
 \left[
\frac{I}{0.405 \pi h }
\right]^ \frac{1}{n+1}
\frac{\mathbf{b} \rho^{cr}}{8} \lambda \\
\end{equation}
The $\tilde{\epsilon}_E(\theta) $ values are not generally known and (as with \cite{LiuZhang2004}) the bulk failure stress has to be measured relative to a reference state where the particles are too widely spaced to affect the ductility.

\begin{figure} [htpb] 
\begin{center}
\includegraphics{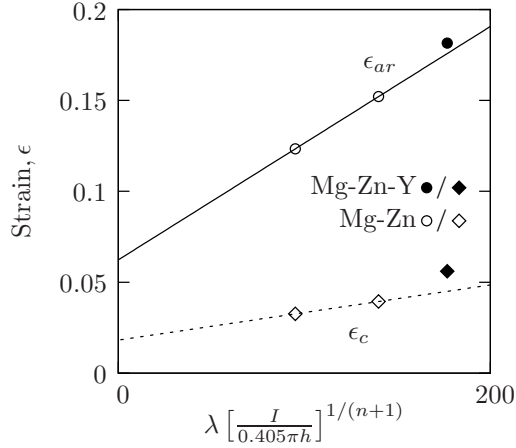}
\caption{\label{fig-DuctilityWh}
Tensile ductility as a function of \betap precipitate distribution. 
The true strain at  failure (by area reduction) ($\epsilon_{ar}$) and true strain onset of instability (i.e. uniform elongation, $\epsilon_c$) are plotted as a function of the particle spacing multiplied by Ramberg-Osgood work hardening factor.
Open symbols indicate values for basal slip and filled symbols indicate prismatic slip. 
Dashed lines indicate least squares fit for uniform elongation (diamonds), while solid lines show a least squares fit for true strain at failure (circles).
}
\end{center}
\end{figure}

The uniform ductility and true strain at failure are plotted as a function of the  precipitate spacing multiplied by the work-hardening correction, i.e. $ \left(  \lambda \left[
\frac{I}{0.405 \pi l }
\right]^ \frac{1}{n+1} \right)$ in Figure~\ref{fig-DuctilityWh}.
The true strain at the necked region was calculated from the area reduction of the failed samples and is shown as $\epsilon_{ar}$. 
The uniform elongation was calculated from the tensile curves using the Consid\'ere criterion, describing the commencement of necking where the true strain exceeds the strain-hardening rate, i.e. 
\( \sigma \ge \left( {\frac{d\sigma}{d\epsilon} } \right) \label{eqn-considere} \).

For particle spacings calculated by basal slip (open symbols), $\epsilon_{ar}$(indicated by diamonds) and $\epsilon_{c}$ (circles)  increased approximately linearly with the particle spacing at similar rates ($\sim1.6\times10^{-3}$\,nm$^{-1}$) as expected from Equation~\ref{eqn-critical-stress}. 
Similarly, for prismatic slip  (filled symbols), the rate of increase of $\epsilon_{ar}$(diamonds) and $\epsilon_{c}$ (circles)  were similar one another, but at at much lower rate of $\sim 9\times10^{-5}$\,nm$^{-1}$. 
If  basal slip was predominant this would suggest that precipitates with a $\lambda$ spacing greater than approximately 200\,nm would have negligible effect on the ductility. 
This may well underestimate the effect of \betap precipitates on ductility.
For prismatic slip, particle spacings up to $1\mu$m would have an effect on the ductility.
A more detailed study of the precipitate-dislocation interaction is currently underway.

\subsection{The effectiveness of pre-ageing deformation}

Aside from the pioneering study by Clark \cite{ClarkMgZn1965} there has been very limited quantitative work on the precipitation strengthening response in Mg-Zn alloys. 
Although this early work was valuable in establishing that there was a strong precipitation hardening response, the lack of control over the deformation poses difficulties in understanding the effectiveness of pre-ageing deformation in enhancing \betap nucleation. 
In Clark's study, pre-ageing deformation was carried out by cold-rolling Mg-Zn sheet; however, this would inevitably result in extensive deformation twinning in addition to dislocation glide. 
The relative extent of twinning and slip will depend on the orientation of each individual grains and  it is likely that the precipitation-strengthening response will vary between grains. 
This makes it impossible to properly connect the microstructure to the bulk mechanical properties. 

In the present study, the use of  texture combined with controlled tensile deformation ensured that twinning was avoided and that changes in the precipitation response can be ascribed unambiguously to nucleation on dislocations. 
This is particularly important since twins also act as strong nucleation sites for coarse (and therefore poorly strengthening) 
\betap and \ce{MgZn2} particles. 
Since the amount of deformation was closely controlled, it was possible to determine the net plastic strain, rather than the 
total applied strain. 
This allowed the elastic strain (which does not contribute to dislocation multiplication) to be excluded from consideration. 

The majority of the the improvement in the yield strength was achieved with 3\% plastic strain with the response diminishing  at higher strains. 
The ageing temperature of 150$^\circ$ is within the temperature range used for stress annealing of Mg-Zn alloys \cite{AvedesianMg99} and the diminishing return is due to the annealing out of deformations.
This is evident in the hardness test results (Fig~\ref{fig-hardness}) that show an increase in hardness after pre-ageing deformation (indicating work-hardening) and an immediate drop in hardness upon ageing as the stress is relieved. 
A similar effective was noted by Clark \cite{ClarkMgZn1965} for 10 and 20\% rolling reductions, but as noted above the complex nature of the deformation made the observation difficult to interpret.
The rapid annealing out of dislocations indicates that the effective window for pre-ageing deformation is quite narrow and that little further improvement in the strength is to be expected for purely tensile deformation of more than 5\% strain. 

\subsection{The effect of inhomogeneous precipitation}

The inhomogeneity of the precipitate distribution shows a substantial effect on the strength and ductility, but one that is rarely given serious consideration. 
It is well known that the effectiveness of the precipitates in restricting dislocation motion depends on the inverse spacing between precipitates.
However, how this spacing is calculated and expressed often escapes notice.
The standard approach of determining an average spacing from the volume fraction and number density does not take any account that the precipitate distribution is inhomogeneous. 
The Delaunay triangulation method applied in the present work provides an average value of the interparticle spacing that takes into account inhomogeneity in the particle distribution. 
Along with related methods such as Voronoi decomposition \cite{EstrinVoronoi1999} it has been shown to provide a more realistic measure of the effective particle spacing \cite{Lepinoux2000, Lepinoux2001}.
This work has demonstrated that this calculated $\lambda$ values underestimates the effective spacing between particles, resulting in an overestimate of the Orowan strengthening effect. 

\begin{figure} [htpb] 
\begin{center}
\includegraphics{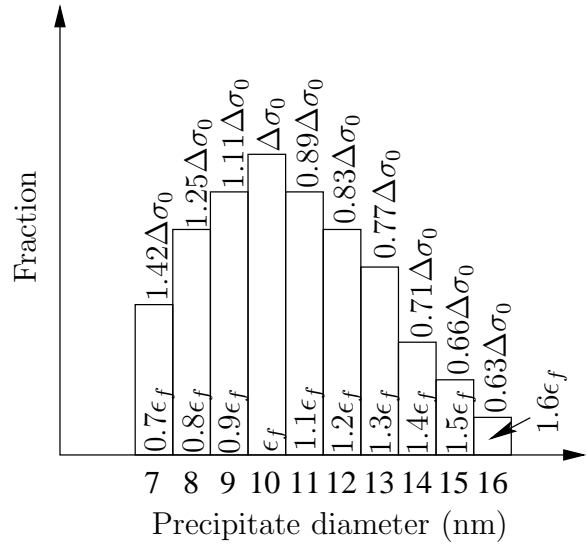}
\caption{Schematic of the effect of the spread of a hypothetical particle diameter distribution on yield strength and ductility.
The yield strength increment ($\Delta\sigma$) and elongation to failure ($\epsilon_f$) are shown relative to the value for the  mode of the distribution (10\,nm). 
\label{fig-Spread}}
\end{center}
\end{figure}

In addition to the variation between measured and calculated $\lambda$ values, it is also important to consider the effect of the spread of particle spacings on the mechanical properties. 
Unfortunately,  the individual spacings between precipitates were not directly accessible with the current software.
However, the diameter distributions are known and as a first approximation the system can be divided into subsystems of arbitrary width and number fraction, each having a homogeneous diameter value.
Each subsystem will be characterised by a given particle diameter and spacing and hence yield strength and ductility  (Shown schematically in Fig~\ref{fig-Spread}).

The  precipitate diameter distributions (Fig~\ref{fig-diameter}) all showed positive skew (that is, the majority of the diameter values were less than the mean) and it is reasonable to infer that the distribution of particle spacings will therefore show negative skew, (that is, the majority of the $\lambda$ values being greater than the mean).
This indicates that large volumes of the matrix will be poorly strengthened, while smaller regions will be substantially stronger than average. 
This is consistent with the TEM observations of clusters of particles along dislocations  and large,  regions with low particle densities and appears reasonable for a  microstructure where nucleation occurs preferentially at heterogeneous sites. 
Such a negatively skewed precipitate spacing is particularly undesirable for mechanical properties. 
The large volumes of poorly strengthened matrix will experience slip at lower stresses than the average spacing would indicate, while misfit dislocations will rapidly build up  in regions with closely-spaced particles, leading to premature failure. 
However, the effect on yield strength and ductility are subtly different, due to the work hardening behaviour.

The spread of yield strength values from lowest (high particle spacing) to highest (low particle spacing) means that slip will occur first where the particles are most widely dispersed; i.e. large volumes of the matrix in which precipitates are sparse, gliding at low flow stresses until they intersect with a precipitate.  
However, when the particles are resistant to shearing by dislocations, dislocation loops will either cross-slip or accumulate around the bypassed precipitates, resulting in work-hardening. 
This feedback will impede further glide through those regions of the matrix, effectively reducing the disparity in yield strength between  regions of the matrix. 

The situation with ductility is somewhat different.
Ductile failure is assumed to take place when the dislocation density around a given particle reaches a critical threshold,
which is determined by the particle spacing.
The alloy can again be divided into subsystems of different spacings and hence critical imposed strain.
Dislocations accumulate most rapidly in regions where the particles are closely spaced, however, unlike the strengthening behaviour there is no effective feedback.
Once necking commences in these regions it will accelerate rather than impede failure in the surrounding volume.

It might be expected, therefore that precipitation strengthening--particularly  where there is a broad range of particle spacings--would have a more pronounced effect on ductility than on strength.
This appears to be borne out in the experimental results where, comparing the non-deformed and 5\% strained samples the strength increases by $\sim$110\%, while the ductility was reduced to 35\% of the original value.  

\section{Conclusions}
The size and spacing of rod-like \betap precipitates in a Mg-Zn alloy as been modified by pre-ageing deformation (0\%, 3\% and 5\%). 
A quantitative evaluation was made of the effect of the diameter ($d$), length ($l$) and inter-precipitate spacing on the basal planes ($\lambda$) on the strength and ductility of the alloys. 
To avoid complications due to possible twinning during pre-ageing deformation, texture was imparted to the alloys by extrusion. 
The following conclusions were drawn.

\begin{itemize}

\item Length of the precipitates in the peak-aged alloys decreased from 440 nm for ageing without pre-ageing deformation to 60 nm for 5 \% pre-ageing deformation. 
Correspondingly, the average diameter of the precipitates decreased from 14 nm to 9 nm. 

\item The precipitate distribution was somewhat inhomogeneous, resulting in the interparticle spacing on the basal plane being approximately 15\% greater than for a fully homogeneous case. 
It was considered probable that interparticle spacings on the prismatic planes are similarly underestimated by the Orowan equations.

\item The yield strength ($\sigma_y$) increased from 273\,MPa (no strain) to 309\,MPa (5\% strain). 
There was a rapidly diminishing return for the increase of yield strength with increasing deformation, with 3\% strain being nearly as effective as 5\%. 
The increase in yield strength was roughly linearly proportional to the interparticle spacing on basal or prismatic planes, however, the rate of increase (basal: 3100\,MPa.nm, prismatic 3550\,MPa.nm) was lower than calculated for Orowan looping of rod-shaped particles (4760\,MPa.nm). 

\item The ductility (tensile elongation to failure ) decreased from 17\% to 6\%.
The true fracture strain and uniform elongation showed a linear relationship with the precipitate spacing on basal and prismatic planes, consistent with models for ductile failure due to the accumulation of geometrically necessary dislocations around the precipitates. 
The model predicts a much more rapid reduction in ductility with particle spacing on basal slip, compared to prismatic slip.
Investigations are underway to clarify which mode predominates during deformation of alloys containing closely-spaced particles.

\end{itemize}

\section*{Acknowledgements}

One of the authors (JMR) gratefully acknowledges the support of the Japan Society for the Promotion of Science (JSPS) through a JSPS fellowship. 
The authors also thank  Reiko Komatsu, Keiko Sugimoto and Toshiyuki Murao for assistance with sample preparation.

\end{document}